\documentclass[10pt,leqno,twoside]{article}
\usepackage[utf8]{inputenc}
\usepackage{mathrsfs,amssymb,amsfonts,amsthm,amsmath,amsbsy}
\usepackage{dsfont,verbatim,fancyhdr}
\usepackage{enumerate}
\usepackage{marginnote}
\usepackage{graphicx,float}
\usepackage{cancel}
\usepackage[hang]{caption}
\usepackage[usenames,dvipsnames]{xcolor}
\usepackage[bookmarks,bookmarksnumbered,colorlinks=true,pdfstartview=FitV, linkcolor=Red,citecolor=blue!90!green,urlcolor=blue!90!green]{hyperref}
\usepackage{tikz}
\usepgflibrary{arrows}
\usepackage{bm}
\usepackage{multirow}
\usepackage{authblk}
\usepackage{framed} 
\usepackage{makecell}
\usepackage{array}
\usepackage{natbib}
\usepackage{diagbox}

\allowdisplaybreaks
\definecolor{ired}{rgb}{0.9,0,0.1}

\def\L{\mathcal{L}}

\def\P{\mathbb{P}}
\def\E{\mathbb{E}}
\renewcommand{\l}{\left}
\renewcommand{\r}{\right}

\newtheorem{theorem}{Theorem}[section]
\newtheorem{remark1}[theorem]{Remark} \newenvironment{remark}{\begin{remark1}\rm}{\hfill$\square$\end{remark1}}
\newtheorem{assumption}{Assumption}[section]

\newcolumntype{P}[1]{>{\centering\arraybackslash}p{#1}}
\newcolumntype{M}[1]{>{\centering\arraybackslash}m{#1}}

\makeatletter
\def\keywords{\xdef\@thefnmark{}\@footnotetext}
\makeatother

\title{\bf \vspace{-2.5cm}
A Test for Differential Ascertainment in Case-Control Studies with Application to Child Maltreatment}

\author{\Large{\textsc{Matteo Sordello and Dylan S. Small}} \\
\large{\textit{Department of Statistics, Wharton School, University of Pennsylvania}}}
\date{}

\begin{document}
\thispagestyle{empty}

\maketitle

\begin{center}
\begin{minipage}{.75\textwidth}
\footnotesize\noindent

We propose a method to test for the presence of differential ascertainment in case-control studies, when data are collected by multiple sources. We show that, when differential ascertainment is present, the use of only the observed cases leads to severe bias in the computation of the odds ratio. We can alleviate the effect of such bias using the estimates that our method of testing for differential ascertainment naturally provides.
We apply it to a dataset obtained from the National Violent Death Reporting System, with the goal of checking for the presence of differential ascertainment by race in the count of deaths caused by child maltreatment.

\end{minipage}
\end{center}
\linespread{1.2}


\section{The Problem of Differential Ascertainment in Case-Control Studies}

In case-control studies, one wants to understand the association between some risk factor and a rare outcome. 
The odds of a risk factor are compared among cases (subjects with the outcome) vs. controls (subjects without the outcome).  When it is possible to get a complete count of the cases or to obtain a random sample of the cases, then estimating the prevalence of the risk factor among the cases is straightforward. But for some settings, it is not possible to get a complete count or take a random sample of the cases.  This problem does not arise strictly in case-control studies, but in any situation in which one wants to count the number of individuals that belong to a certain group. In such circumstances we say that we are in the presence of \textit{imperfect ascertainment}. It can happen in a multitude of different contexts, for example when counting the number of people addicted to a certain drug \citep{dhmsa95}, civilians that died during a war \citep{bbsda99, pgb14} or people that suffer from diabetes \citep{iwg95}. A further complication might arise if the group of people under consideration is divided into two subpopulations, according to a variable. In such cases, the ascertainment can be not only imperfect, but possibly easier in one of these two subpopulations, and harder in the other. For example, one can think that there is a threshold on the variable age under which it is easier to ascertain if a person is a drug addict, or that gender is a relevant factor in the ascertainment of civilian deaths or of individuals suffering from diabetes. In practical terms, this means that the ratio of the observed counts in the two subpopulations is different from the real underlying ratio. Here we say that we are in the presence of \textit{differential ascertainment}, a situation that is harder to detect than simple imperfect ascertainment.

In this work, we focus on a specific situation where we suspect that differential ascertainment could be involved, and develop the machinery to perform a rigorous statistical analysis.
The question we want to answer concerns deaths caused by maltreatment of children aged 10 or less. Notice that here we are in the presence of imperfect ascertainment, since it is not always possible to assess whether or not a death has been caused by child maltreatment \citep{hbvchhb99}.
We are interested in knowing whether the deaths caused by child maltreatment are differentially ascertained according to the variable race. There are two main reasons we are interested in this question. The first one is that most of the research done in this area has the goal of providing guidelines that can be used to develop and evaluate intervention strategies, ultimately to reduce the number of deaths caused by child maltreatment. If the real proportions in the population are different from the ones observed, then the resources put forward to solve the problem may be poorly allocated and less effective than they should be.
The second reason is that differential ascertainment can lead case-control studies to provide false conclusions. To explain this, we introduce some notation for case-control studies (for a thorough discussion of case-control studies see \citet{kc14}). For a case-control study, let $e \in \{E, U\}$ be the bivariate exposure, $E$ stands for \textit{exposed} and $U$ for \textit{unexposed}, and $Y \in \{0, 1\}$ be the outcome of interest, where $Y = 1$ is the indicator of the cases and $Y = 0$ of the controls. We would like to infer the distribution of $Y | e$, and in particular to compute the risk ratio (RR), which is defined as $\P(Y=1|e=E)/\P(Y=1|e=U)$. Unfortunately, given that it is structured retrospectively, a case-control study can only provide information on the distribution of $e | Y$. The following equality comes to our aid, creating a link between these distributions.
\begin{equation*}
\frac{\P(e = E | Y = 1)}{\P(e = U | Y = 1)}\cdot \frac{\P(e = U | Y = 0)}{\P(e = E | Y = 0)} = \frac{\P(Y = 1 | e = E)}{\P(Y = 0 | e = E)}\cdot \frac{\P(Y = 0 | e = U)}{\P(Y = 1 | e = U)} . \\[5pt]
\end{equation*}
The left hand side is called the odds ratio (OR), and it is a fundamental quantity to assess the association between exposure and outcome. Let $T^E$ be the total number of exposed individuals and $T^U$ the total number of unexposed individuals in the whole population of interest (so counting both cases and controls). We assume that $T^E$ and $T^U$ are known. Now let $N^E$ be the number of exposed cases and $N^U$ the number of unexposed cases.
The odds ratio is then
\begin{equation}{\label{or_equation}}
OR = \frac{\frac{N^E}{N^E + N^U}\cdot \frac{T^U - N^U}{T^U + T^E - N^U - N^E}}{\frac{N^U}{N^E + N^U}\cdot \frac{T^E - N^E}{T^U + T^E - N^U - N^E}} = \frac{N^E \cdot (T^U - N^U)}{N^U\cdot (T^E - N^E)} . \\[3pt]
\end{equation}
We assume throughout the paper that it is possible to obtain a random sample from the two groups of exposed and unexposed controls, or simply that the ratio $(T^U-N^U)/(T^E-N^E)$ is known or can be estimated.
Plugging the estimate of the ratio $(T^U-N^U)/(T^E-N^E)$ into the odds ratio (\ref{or_equation}) reduces the problem of estimating OR to estimating the ratio between the exposed and unexposed cases, $N^E/N^U$. The distribution of OR will depend on the randomness in the estimates for $N^E/N^U$ and $(T^U-N^U)/(T^E-N^E)$ combined.

If there is imperfect ascertainment of the cases, the real values of $N^E$ and $N^U$ are unknown.
If one makes the assumption that the real proportion of exposed to unexposed cases is consistently estimated by the observed proportion, then plugging the observed proportion into (\ref{or_equation}) provides a consistent estimate of the OR.  However, if there is differential ascertainment, we know by definition that the use of the observed ratio instead of the real $N^E/N^U$ will lead to some bias in the computation of OR, which can be severe, as we will show in Section \ref{effect_differential_ascertainment}.

Another situation where we would care about the count of exposed and unexposed cases is in a case-cohort study \citep{pre86}, in which the cases have imperfect ascertainment but the controls have perfect ascertainment. In case-cohort studies, a census of cases in a cohort is taken and then a sub-cohort of the cohort is sampled (the sample includes both cases and control). For a case-cohort study with perfect ascertainment of cases, one can estimate the risk ratio by observing that
\begin{equation*}
RR = \frac{\P(Y=1|e=E)}{\P(Y=1|e=U)} = \frac{\P(e=E | Y=1) \cdot \P(e=U)}{\P(e=U | Y=1)\cdot \P(e=E)} = \frac{N^E\cdot T^U}{N^U\cdot T^E} ; \\[3pt]
\end{equation*} 
the quantity $N^E/N^U$ is obtained from the census of cases and $T^U/T^E$ is estimated from the sub-cohort which is a random sample of the population in the cohort. If there is imperfect ascertainment of cases, then assuming that the cases are only a small fraction of the population in the cohort, the exposure ratio in the population, $T^U/T^E$, can still be estimated from the sub-cohort and again our attention focuses on finding an estimate for $N^E/N^U$.

Problems of this nature, where the total size of a population is unknown, have been deeply studied in the capture-recapture literature. 
\citet{bfh75} and \citet{w82} wrote extensively on incomplete tables and how to estimate the size of a closed population. \citet{c89} considered various extensions, including the situation where the target population is open.
\citet{bvmfphsh16} compared different capture-recapture estimators in situations where their assumptions are violated, while \citet{shs00} and \citet{lshh01} discussed the problem of matches between lists that are not recorded correctly.
One commonly used model for the catchability of an individual in the capture-recapture literature is the Rasch model \citep{r60, a94}.
Here we consider a simple modification of the original Rasch model.  The model results in $N^E$ and $N^U$ being realizations of a Poisson random variables with means $\gamma_E$ and $\gamma_U$ respectively. 
We estimate those means, and introduce the parameters $\theta_j$ representing the amount of differential ascertainment in each list. 
We use these $\theta_j$'s to test if the differential ascertainment is significantly different from zero. 
If it is significant, then we use $\hat{\gamma}_E$ and $\hat{\gamma}_U$ in the computation of OR or RR.
If it is not significant, then the observed counts of exposed and unexposed cases can be used instead of the estimated means.

\section{The Model}{\label{model}}

As anticipated before, we consider a bivariate exposure $e \in \{E, U\}$ and a total number of cases $N^e$ in each group. We first introduce the classical version of the Rasch model \citep{r60}, which relies on two important assumptions.
\begin{assumption}[Individual Independence]{\label{individual_independence}}
The ascertainment of each individual in any list is independent from the ascertainment of any other individual.
\end{assumption}
\begin{assumption}[List Independence]{\label{list_independence}}
For any individual, the ascertainment happens in a list independently from any other list.
\end{assumption}
\noindent
Assumption \ref{list_independence} is not completely reasonable, since we can imagine that being ascertained in one or more lists makes an individual more or less prone to be ascertained again. 
This mechanism is explained in \citet{bfh75} under the name of social visibility/invisibility or trap fascination/avoidance, depending on the context. In Section \ref{list_dependence} we will get rid of this assumption and explain how to include dependence among the lists. 
We will sometimes refer to the concept of ascertainment as capture, given that our framework is essentially the same as the capture-recapture one.

The number of lists used is denoted by $J$.
Conditionally on the value $N^e$, the ascertainment matrix $X^e$ has dimension $N^e \times J$, and, for $i = 1,..., N^e$ and $j = 1,..., J$, each entry takes values
$$X^e_{i j} = \l\{\begin{array}{ll}
1 \qquad &\mbox{if individual $i$ is ascertained by list $j$}\\
0 & \mbox{otherwise} \end{array}\r.$$
These Bernoulli random variables take value $1$ with probability
\begin{equation}{\label{rasch_prob}}
\P(X^e_{ij} = 1) = \frac{e^{\theta^e_{ij} + \alpha_j}}{1 + e^{\theta^e_{ij} + \alpha_j}}
\end{equation}
The parameters $\theta^e_{ij}$ represent the individual propensity for an individual with exposure $e$ to be captured from list $j$, and $\alpha = (\alpha_1, ..., \alpha_J)$ are the capture strengths of each list, which are assumed to be independent from the exposure. 
This model naturally creates, for each exposure $e$, a contingency table $M^e$ with $2^J$ entries, each entry being the number of individuals that are ascertained in a specific combination of lists. We denote $M^e_{k_1,..., k_J}$, where $k_1,..., k_J \in \{0, 1\}$, to be the count of individuals that are ascertained in all the lists with corresponding index that takes value $1$, and not ascertained in the lists with index $0$. 
For example, when $J = 3$ (the setting we will use from now on, since it is what we will be dealing with in the NVDRS dataset), $M^e_{101}$ is the count of individuals that appear in list $1$ and list $3$, but not in list $2$. In Table \ref{contingency_table_3lists} we show a contingency table for the three lists setting.

\begin{table}
\centering
\caption{Complete contingency table $M^e$ for the count of individuals with exposure $e$ in the three-list setting.}
\label{contingency_table_3lists}
\vspace{5mm}
\begin{tabular}{M{1.3cm} M{0.7cm} M{1.2cm} M{1.2cm} M{1.2cm} M{1.2cm} @{}m{0pt}@{}}
\cline{3-6}
 &  & \multicolumn{4}{c}{List 1} & \\[5pt] \cline{3-6} 
 &  & \multicolumn{2}{c}{Y} & \multicolumn{2}{c}{N} & \\[5pt] \cline{3-6} 
 &  & \multicolumn{4}{c}{List 2} & \\[5pt] \cline{3-6} 
\multicolumn{1}{c}{} &  & Y & N & Y & N & \\[5pt] \cline{1-6} \\[-10pt]
\multicolumn{1}{c}{\multirow{2}{*}{List 3}} & \multicolumn{1}{c}{Y} & $M^e_{111}$ & $M^e_{101}$ & $M^e_{011}$ & $M^e_{001}$ & \\[5pt] \cline{2-6} \\[-10pt]
 \multicolumn{1}{c}{}&\multicolumn{1}{c}{N} & $M^e_{110}$ & $M^e_{100}$ & $M^e_{010}$ & $M^e_{000}$ & \\[5pt] \cline{1-6} 
\end{tabular}
\end{table}
We consider two simplified versions of this model. In one we assume that the value of the individual capture propensity is zero for all the unexposed cases, and only depends on the list, but not the specific individual, for the exposed. In this framework we set $\theta^U_{ij} = 0$ and $\theta^E_{ij} = \theta_j$. Then we further simplify the model by only allowing a single parameter $\theta$ for all the exposed cases and all the lists. While the former setting allows more flexibility in the model, the second one provides a clear answer to the question of the presence of differential ascertainment in the data (see Remark \ref{why_theta}).
We will explain how to test if it is necessary to consider different values for each $\theta_j$. The following description of the model will use the notation where the capture propensity is list dependent for the exposed cases.

In the Appendix we introduce another variation, where we allow each individual to have a different propensity to be captured. We make the assumptions that exposed cases have individual propensity $\theta_{ij}^E \sim N(\mu, \sigma^2)$, and unexposed individuals have $\theta_{ij}^U \sim N(0, \sigma^2)$. The magnitude of the newly introduced parameter $\sigma$ can tell us if there is significant variability among the individuals.
We will see, however, that in the study of the NVDRS data we don't really need to introduce this variability, since our estimate of $\sigma$ is going to be extremely small and all the other parameters very close to the ones estimated with the initial setting.

Using Assumption \ref{list_independence}, we compute the probability for each exposed individual to appear in a specific cell identified by subscripts $x y z$, which is just the product of the single lists' probabilities
\begin{equation}{\label{prob_rasch_indep_exp}}
p^{E}_{xyz} := \P\l(X^E_{i1} = x, X^E_{i2} = y, X^E_{i3} = z\r) = \frac{e^{x\cdot(\theta_1 + \alpha_1)}}{1 + e^{\theta_1 + \alpha_1}} \cdot \frac{e^{y\cdot(\theta_2 + \alpha_2)}}{1 + e^{\theta_2 + \alpha_2}} \cdot \frac{e^{z\cdot(\theta_3 + \alpha_3)}}{1 + e^{\theta_3 + \alpha_3}}
\end{equation}
while for unexposed cases we have
\begin{equation}{\label{prob_rasch_indep_unexp}}
p^{U}_{xyz} := \P\l(X^U_{i1} = x, X^U_{i2} = y, X^U_{i3} = z\r) = \frac{e^{x\cdot\alpha_1}}{1 + e^{\alpha_1}} \cdot \frac{e^{y\cdot\alpha_2}}{1 + e^{\alpha_2}} \cdot \frac{e^{z\cdot\alpha_3}}{1 + e^{\alpha_3}}
\end{equation}
Thanks to Assumption \ref{individual_independence}, the likelihood of a complete table $M^e$ with total count $N^e$ is given by a multinomial distribution
\begin{equation}{\label{multinomial_likelihood}}
\L(M^e | N^e) \sim \text{Multinomial}(M^e | N^e,\ p_{xyz}^e)
\end{equation}
where the probabilities are given by (\ref{prob_rasch_indep_exp}) or (\ref{prob_rasch_indep_unexp}). A consequence of Assumption \ref{individual_independence} is that the contingency tables associated with different exposures are independent. Then the total likelihood of the two complete contingency tables we are considering is
\begin{equation}{\label{complete_model}}
\L(M^E, M^U | N^E, N^U) = \L(M^E | N^E)\cdot\L(M^U | N^U) .
\end{equation}
Notice that up to now we were able to condition on the total number of cases $N^e$, since we were dealing with contingency tables where no cell was missing. In real applications, however, the contingency tables are not complete, and the cells $M^e_{000}$ are missing, since they represent the elements that are not ascertained in any list. This means that $N^e$ is now unknown, and the definition of the likelihood is not as immediate as before.
We then add the assumption that, for any $e$, the count $N^e \sim \text{Poisson}(\gamma_e)$.
If we let $M_{obs}^e$ be the contingency table without the unobserved cell and $N_{obs}^e$ the corresponding count of all the observed cases, then the likelihood of the observed table is 
\begin{align}{\label{multinomial_likelihood_observed}} \nonumber
\L(M^e_{obs}) &= \sum_{m = 0}^\infty \L(M^e_{obs}, M^e_{000} = m) \\
&= \sum_{m = 0}^\infty \L(M^e_{obs}, M^e_{000} = m, N^e = N^e_{obs} + m) \\ \nonumber
&= \sum_{m = 0}^\infty \L(M^e_{obs}, M^e_{000} = m | N^e = N^e_{obs} + m) \cdot \P(N^e = N^e_{obs} + m)
\end{align}
where $\l\{M^e_{obs}, M^e_{000} = m | N^e = N^e_{obs} + m\r\}$ is a complete contingency table with $m$ observations in the bottom-right cell, hence distributed according to a multinomial distribution with probabilities given in (\ref{prob_rasch_indep_exp}) or (\ref{prob_rasch_indep_unexp}), and $N^e$ is Poisson with mean $\gamma_e$.
In its extended form, the likelihood of one incomplete contingency table is then
\begin{align}{\label{extended_likelihood}} \nonumber
\L(M^e_{obs}) &= \sum_{m=0}^\infty \l[ \frac{(N_{obs}^e + m)!}{m! \cdot \prod_{x,y,z} M_{xyz}^e!} \cdot {(p_{000}^e)}^m \cdot \prod_{x, y, z} {(p_{xyz}^e)}^{M_{xyz}^e}\r] \cdot \l[\frac{\gamma_e^{N_{obs}^e + m} e^{-\gamma_e}}{(N_{obs}^e + m)!}\r] \\ \nonumber
&= \frac{\prod_{x,y,z} {(p_{xyz}^e)}^{M_{xyz}^e} \cdot \gamma_e^{N_{obs}^e}\cdot e^{-\gamma_e}}{\prod_{x,y,z} M_{xyz}^e!} \cdot \sum_{m=0}^\infty \frac{(\gamma_e\cdot p_{000}^e)^m}{m!} \\
&= \frac{\prod_{x,y,z} {(p_{xyz}^e)}^{M_{xyz}^e} \cdot \gamma_e^{N_{obs}^e}\cdot e^{\gamma_e (p_{000}^e - 1)}}{\prod_{x,y,z} M_{xyz}^e!}
\end{align}
where the products are only on the $x, y, z$ such that $x + y + z > 0$. The likelihood of the two incomplete contingency tables is, again using Assumption \ref{individual_independence},
\begin{equation}{\label{incomplete_model}}
\L(M^E_{obs}, M^U_{obs}) = \L(M^E_{obs}) \cdot \L(M^U_{obs}).
\end{equation}

\subsection{Dependence among lists}
{\label{list_dependence}}

While the independence among the individuals is an important assumption, we can extend this model to include dependence among the lists. We consider the dynamical Rasch model \citep{vg93}, which eliminates the independence assumption among the lists with a simple modification of the probabilities in (\ref{rasch_prob}). Instead of having them fixed, we generate them sequentially based on the previous realizations of the ascertainment variables. These are the same for each individual, and are defined, for those that are exposed, as
\begin{align*}
p_1^E :=& \P(X^E_{i1} = 1) = \frac{e^{\theta_1 + \alpha_1}}{1 + e^{\theta_1 + \alpha_1}} \\[3pt] 
\Rightarrow \quad &X^E_{i1}\sim \text{Bernoulli}\big(p_1^E\big) \\[6pt] 
p_{2,x}^E :=& \P(X^E_{i2} = 1 | X^E_{i1}=x) = \frac{e^{\theta_2 + \alpha_2 + \alpha_{12}\cdot x}}{1 + e^{\theta_2 + \alpha_2 + \alpha_{12}\cdot x}} \\[3pt] 
\Rightarrow\quad  &X^E_{i2} | X^E_{i1}=x \sim \text{Bernoulli}\big(p_{2,x}^E\big) \\[6pt]
p_{3,x,y}^E :=& \P(X^E_{i3} = 1 | X^E_{i1}=x, X^E_{i2}=y) = \frac{e^{\theta_3 + \alpha_3 + \alpha_{13}\cdot x + \alpha_{23}\cdot y}}{1 + e^{\theta_3  + \alpha_3 + \alpha_{13}\cdot x + \alpha_{23}\cdot y}} \\[3pt]  
\Rightarrow\quad &X^E_{i3} | X^E_{i1}=x, X^E_{i2}=y \sim \text{Bernoulli}\big(p_{3,x,y}^E\big)
\end{align*}
and if the number of lists is greater we can keep generating $\P(X^E_{ik} = 1)$ and $X^E_{ik}$ based on the previous values of $X^E_{ij},\ j = 1,..., k-1$. Here we decided to consider only the two-list interactions, given by the parameters $\alpha_{12}, \alpha_{13}$ and $\alpha_{23}$. It is also possible to use, as the number of lists grows, higher-order interactions.
For unexposed individuals the procedure is analogous, just setting as before $\theta_j = 0$ for $j = 1,2,3$.
Then we have, with the notation introduced before
\begin{equation}
\begin{aligned}{\label{prob_rasch_no_indep}}
p^{E}_{xyz} &= \frac{e^{x\cdot(\theta_1 + \alpha_1)}}{1 + e^{\theta_1 + \alpha_1}} \cdot \frac{e^{y\cdot(\theta_2 + \alpha_2 + \alpha_{12}\cdot x)}}{1 + e^{\theta_2 + \alpha_2 + \alpha_{12}\cdot x}} \cdot \frac{e^{z\cdot(\theta_3 + \alpha_3 + \alpha_{13}\cdot x + \alpha_{23}\cdot y)}}{1 + e^{\theta_3  + \alpha_3 + \alpha_{13}\cdot x + \alpha_{23}\cdot y}} \\[5pt]
p^{U}_{xyz} &= \frac{e^{x\cdot\alpha_1}}{1 + e^{\alpha_1}} \cdot \frac{e^{y\cdot(\alpha_2 + \alpha_{12}\cdot x)}}{1 + e^{\alpha_2 + \alpha_{12}\cdot x}} \cdot \frac{e^{z\cdot(\alpha_3 + \alpha_{13}\cdot x + \alpha_{23}\cdot y)}}{1 + e^{\alpha_3 + \alpha_{13}\cdot x + \alpha_{23}\cdot y}}
\end{aligned}
\end{equation}
The model for two incomplete and independent contingency tables is then the same as before, only with the probabilities in (\ref{prob_rasch_no_indep}) substituted for (\ref{prob_rasch_indep_exp}) and (\ref{prob_rasch_indep_unexp}).
The identifiability of the parameters in this model is discussed in the Appendix.
To fit the model, one simply maximizes the log-likelihood obtained in (\ref{incomplete_model}), where
\begin{equation*}
\log\l(\L(M^e_{obs}) \r) \propto \sum_{x,y,z} M_{xyz}^e \cdot \log\big(\gamma_e\cdot p_{xyz}^e\big) + \gamma_e (p_{000}^e - 1) .
\end{equation*}
A generalized likelihood ratio test can then be performed to verify if it is important to use different values for the $\theta_j$'s in different lists. The null hypothesis is $H_0: \theta_1 = \theta_2 = \theta_3$, and it is tested against the alternative that at least one $\theta_j$ is different from the others. The test statistic that we can use is
\begin{equation}
\Lambda = -2(\ell_0 - \ell)
\end{equation}
where $\ell_0$ is the log-likelihood evaluated in the maximum likelihood estimator computed under $H_0$ (so when all $\theta_j$'s are equal), and $\ell$ is the same estimator found on the model with no constraints. The statistic $\Lambda$ is compared against the quantiles of a $\chi^2$ distribution with $2$ degrees of freedom. If no significant difference is found between the $\theta_j$'s, then the simpler model that only include a single value for $\theta$ can be used.

The next step after obtaining the estimates for the parameters of the model is to check if the estimated ratio $\hat{\gamma}_E/\hat{\gamma}_U$ is substantially different from the observed one, $r:= N^E_{obs}/N^U_{obs}$. Since it is hard to understand if the difference is significant, there are two things that we can do, very similar in practice. Find a confidence interval for the ratio $\hat{\gamma}_E/\hat{\gamma}_U$ or, if we are using the model where a single value for $\theta$ is allowed, perform hypothesis testing on it. In both cases we can use a bootstrap procedure.
\begin{itemize} 
\item To test the null hypothesis $H_0: \theta = 0$, we need to obtain the empirical distribution for the estimate $\hat{\theta}$ under $H_0$. We first estimate the parameters of the model (\ref{incomplete_model}) under the null, and then generate with those parameters new simulated populations; first the new total counts of cases using a Poisson distribution, then the new complete contingency tables using a Multinomial distribution. After excluding the unobserved cells $M^e_{000}$ from the tables, we fit the same model again and obtain the estimate for the parameter $\theta$. This empirical distribution is used to compute the relevant quantiles and check if the point estimate for $\hat{\theta}$ obtained with the real data is significantly far from zero.
\item To get the confidence interval for $\hat{\gamma}_E/\hat{\gamma}_U$ under the null hypothesis $H_0: \gamma_E/\gamma_U = r$, we fit again the model (\ref{incomplete_model}) on the real data imposing the constraint $\gamma_E = r\cdot \gamma_U$. Then, as before, create the new incomplete contingency tables, fit the model and compute and store the ratio of interest. 
\end{itemize}

\begin{framed}
\begin{remark}{\label{why_theta}}
\textbf{Interpretation of the parameter $\theta$ as responsible for differential ascertainment.} Let's consider $N^E$ and $N^U$ fixed. We want to compute $N^E/N^U$ in order to estimate the odds ratio, but since the cells $M^E_{000}$ and $M^U_{000}$ are unobserved we cannot have a precise count of cases. However from (\ref{prob_rasch_no_indep}), when $\theta_1 = \theta_2 = \theta_3 = \theta$ we know that
\begin{align*}
\E\l[M^E_{000}\r] &= N^E \cdot p_{000}^E = \frac{N^E}{(1 + e^{\theta + \alpha_1})\cdot(1 + e^{\theta + \alpha_2})\cdot(1 + e^{\theta + \alpha_3})} \\[5pt]
\E\l[M^U_{000}\r] &= N^U \cdot p_{000}^U = \frac{N^U}{(1 + e^{\alpha_1})\cdot(1 + e^{\alpha_2})\cdot(1 + e^{\alpha_3})}
\end{align*}
We also know that $N^E = \E\l[N_{obs}^E + M^E_{000}\r]$ and $N^U = \E\l[N_{obs}^U + M^U_{000}\r]$. This tells us that $\E\l[N_{obs}^E\r]/\E\l[N_{obs}^U\r] = N^E/N^U$ if and only if $\E\l[M^E_{000}\r]/\E\l[M^U_{000}\r] = N^E/N^U$, which is true only if $\theta = 0$, thanks to the monotonicity of the denominator of $\E\l[M^E_{000}\r]$ with respect to $\theta$. The largest is $|\theta|$, the larger the difference will be between the ratio of the observed counts and the real ratio $N^E/N^U$. Notice also that 
\begin{equation*}
\theta < 0  \quad\Leftrightarrow\quad p^E_{000} > p^U_{000} \quad\Leftrightarrow\quad N^E/N^U > \E[N^E_{obs}]/\E[N^U_{obs}] .
\end{equation*}
If different values for the $\theta_j$'s are allowed, this interpretation is no longer possible. In fact, if one sets for example $\theta_1 = \alpha_2-\alpha_1$, $\theta_2 = \alpha_1-\alpha_2$ and $\theta_3 = 0$, then $p_{000}^E = p_{000}^U$ even though the lists ascertain the cases with different probabilities for exposed and unexposed individuals.
\end{remark}
\end{framed}

\subsection{The Effect of Differential Ascertainment in Estimating OR}
{\label{effect_differential_ascertainment}}

\begin{figure}[]
\caption{The point estimate and $95\%$ confidence bands for the observed ratio $N^E_{obs}/N^U_{obs}$ when using different numbers of lists $J \in \{1, 3, 5\}$. We create differential ascertainment through the parameter $\theta \in [-1, 1]$ and set the main effects to be $\alpha_i = -0.5$.}
\label{confidence_bands_obs_ratio}
\vspace{5mm}
\centering
\includegraphics[width = 0.95\linewidth]{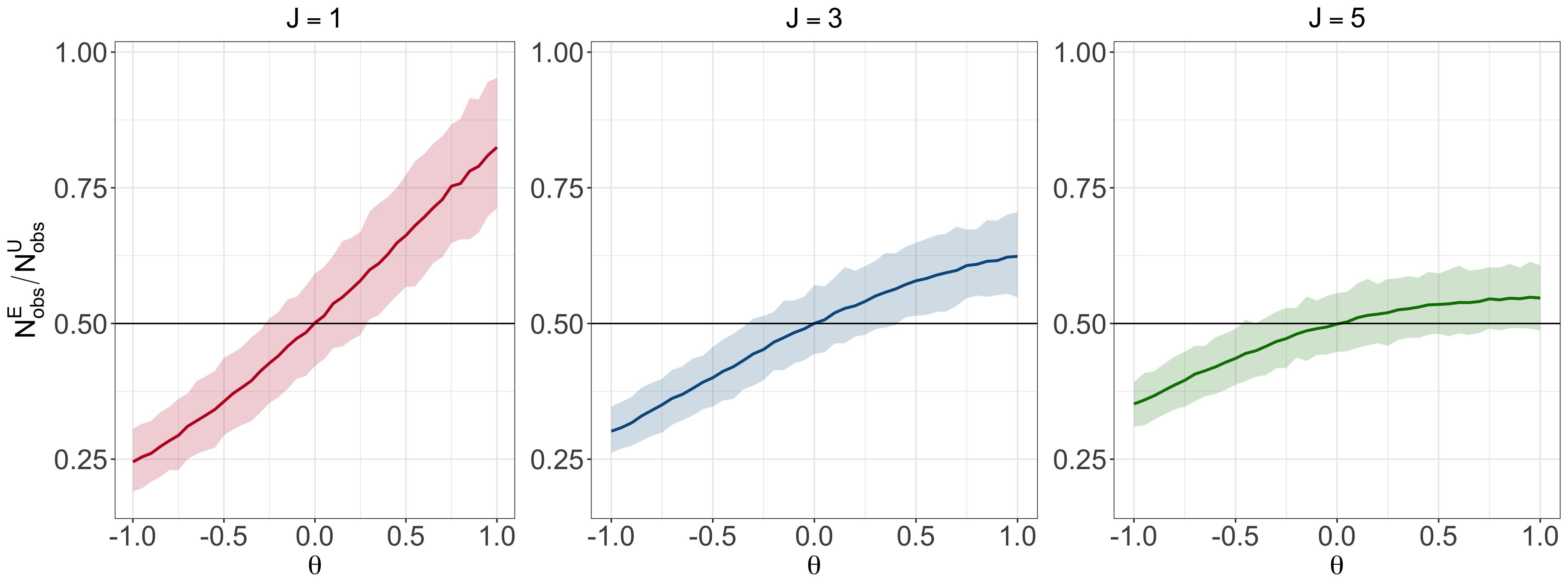}
\end{figure}

We present here a simulation that shows that the use of only the observed counts can lead to severe bias in the estimate of the odds ratio. We set $\gamma_E = 500$ and $\gamma_U = 1000$ to be the average count of cases among the exposed and unexposed population, and consider a single differential ascertainment parameter $\theta \in [-1, 1]$ equal for all lists. We generate the two counts $N^E$ and $N^U$ from a Poisson distribution with mean $\gamma_E$ and $\gamma_U$ respectively.
At every iteration, the ascertainment of the cases happens with the probabilities defined in (\ref{prob_rasch_no_indep}). Here we only use the main effects, set to be $\alpha_i = -0.5$, since the interactions do not influence the counts in the unobserved cells.
We consider the situation where the ascertainment is performed with $J \in \{1,3,5\}$ lists.
The average of the observed ratio of exposed over unexposed individuals is recorded in Figure \ref{confidence_bands_obs_ratio}, together with the $95\%$ confidence bands and the true underlying value $0.5$. The first immediate realization that we have is that, when we created differential ascertainment by setting $\theta \neq 0$, the observed ratio can be extremely biased. It is also clear that the bias increases with $|\theta|$ in each direction (even though non symmetrically), getting far from the true ratio $\gamma_E/\gamma_U = 0.5$. However, such bias gets smaller the more lists we use. This is because, when $J$ grows, even if the individual ascertainment probabilities for each list are small, the chance that at least one list will capture an individual grows, so the count of individuals that are missed by all the lists gets lower. 

This analysis tells us that, in a situation where we don't have a large number of different sources of information available, and we are unsure about the presence of differential ascertainment, it can be risky to just trust the observed ratio. When only one source of information is available, there is unfortunately not much we can do. But we show in the next section how the model just introduced can recognise if this problem if present (estimating the underlying value of $\theta$) and mitigate it using the estimated true counts.

\subsubsection{How the model performs when there is differential ascertainment}
{\label{model_performance}}

\begin{figure}[!htbp]
\caption{The point estimate and $95\%$ confidence bands for the estimated ratio $\hat{\gamma}_E/\hat{\gamma}_U$ (left) and for $\hat{\theta}$ (right) obtained fitting the model (\ref{incomplete_model}) when the differential ascertainment parameter is $\theta \in [-1,1]$. The blue line is the observed mean ratio, reported in the central panel of Figure \ref{confidence_bands_obs_ratio}.
   The data generation process is explained in Section \ref{effect_differential_ascertainment}.}
   \label{confidence_bands_model_ratio}
   \begin{minipage}{0.66\textwidth}
     \centering
     \includegraphics[width=.95\linewidth]{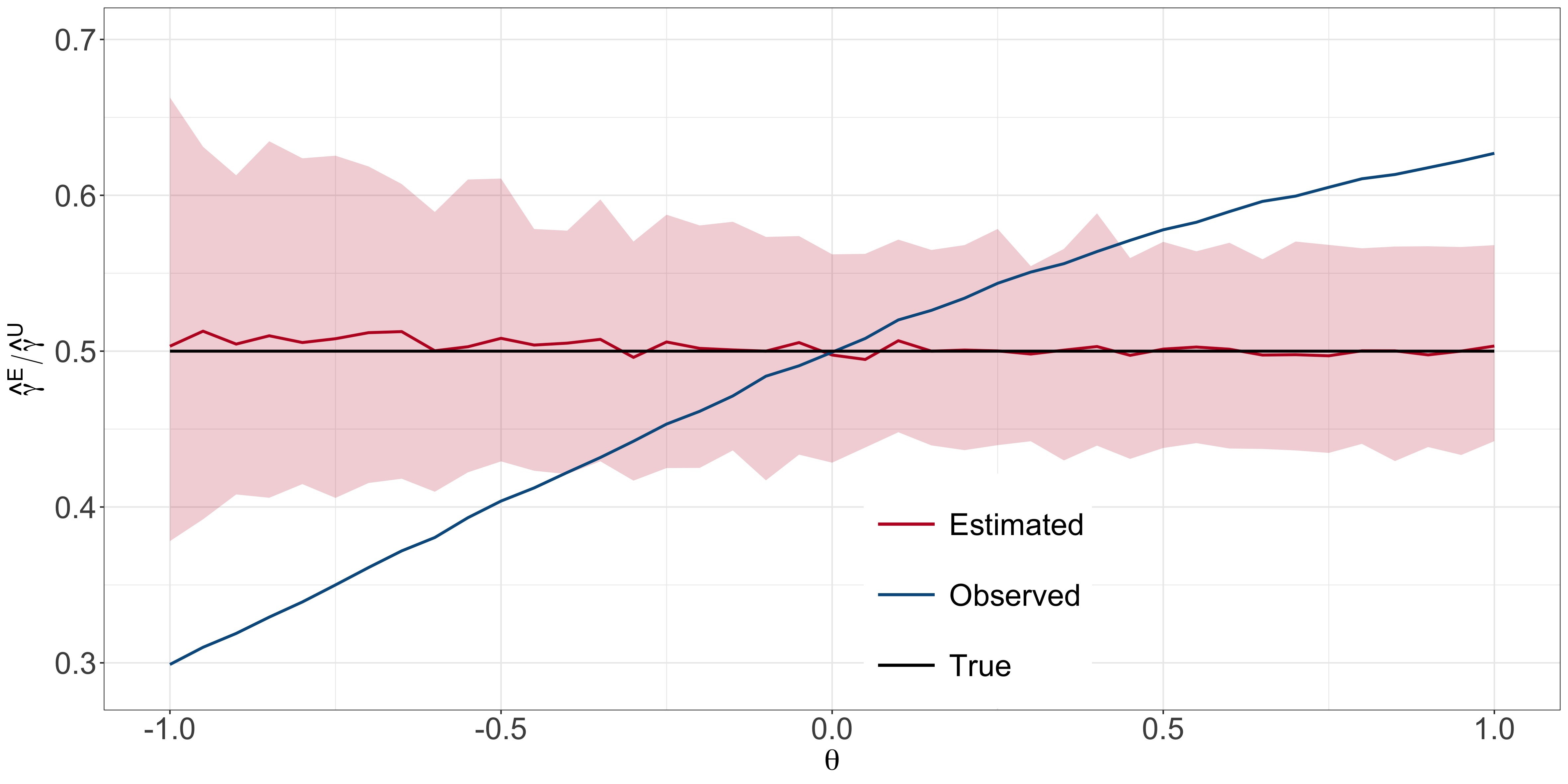}
   \end{minipage}\hfill
   \begin{minipage}{0.33\textwidth}
     \centering
     \includegraphics[width=.95\linewidth]{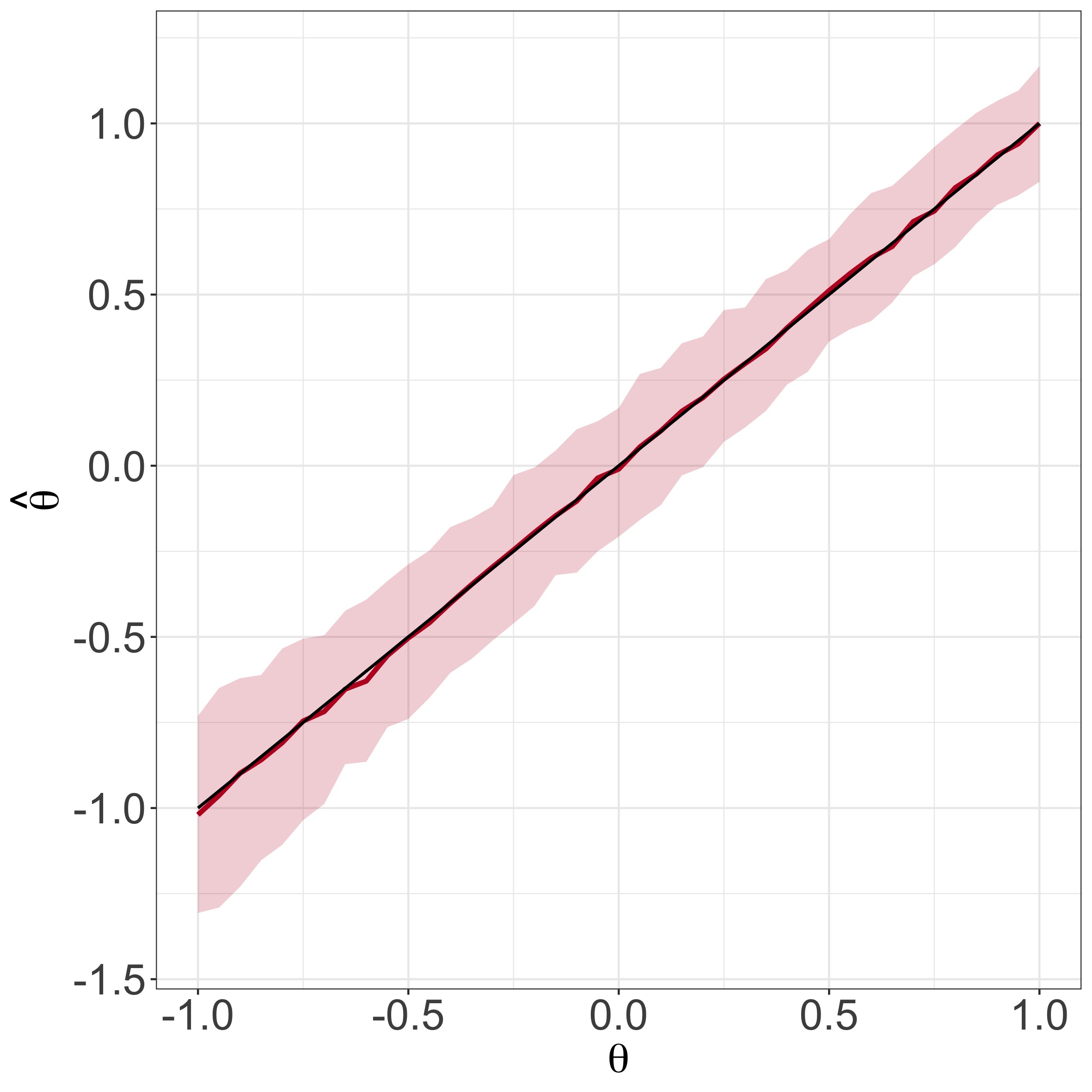}
   \end{minipage}
   \end{figure}

\begin{table}[!htbp]
\centering
\caption{The point estimate and $95\%$ confidence interval for each parameter of the model, when the differential ascertainment parameter is set to be $\theta \in \{-1, -0.5, 0, 0.5, 1\}$. We used $\alpha_i = -0.5$ and $\alpha_{ij} = 0.2$, while the mean counts are $\gamma_E = 500, \gamma_U = 1000$. Each population is generated 200 times and ascertained with 3 lists.}
\label{estimate_parameters_simulation}
\vspace{5mm}
\begin{tabular}{M{.7cm} M{2.2cm} M{2.2cm} M{2.2cm} M{2.2cm} M{2.2cm}@{}m{0pt}@{}}
\hline \\[-10pt]
           & \multicolumn{5}{c}{$\theta$} \\ \cline{2-6} \\[-10pt]
   & -1 & -0.5 & 0 & 0.5 & 1 & \\ \\[-10pt] \hline
 $\hat{\alpha}_1$ & -0.518 \newline (-0.76, -0.28) & -0.507 \newline (-0.74, -0.29) & -0.504 \newline (-0.72, -0.32) & -0.498 \newline (-0.67, -0.32) & -0.504 \newline  (-0.70, -0.32) & \\[9pt] \hline
 $\hat{\alpha}_2$ & -0.522 \newline (-0.84, -0.19) & -0.505 \newline (-0.82, -0.16) & -0.509 \newline (-0.84, -0.22) & -0.508 \newline (-0.78, -0.24) & -0.507 \newline (-0.78, -0.22) & \\[9pt] \hline
 $\hat{\alpha}_3$ & -0.510 \newline (-0.94, -0.08) & -0.504 \newline (-0.91, -0.08) & -0.509 \newline (-0.88, -0.11) & -0.490 \newline (-0.81, -0.13) & -0.490 \newline (-0.88, -0.10) & \\[9pt] \hline
 $\hat{\alpha}_{12}$ & 0.221 \newline (-0.16, 0.59) & 0.203 \newline (-0.14, 0.59) & 0.209 \newline (-0.09, 0.55) & 0.199 \newline (-0.11, 0.51) &  0.213 \newline (-0.08, 0.49) & \\[9pt] \hline
 $\hat{\alpha}_{13}$ & 0.203 \newline (-0.13, 0.56) & 0.193 \newline (-0.13, 0.55) & 0.205 \newline (-0.09, 0.52) & 0.187 \newline (-0.08, 0.50) & 0.183 \newline (-0.09, 0.50) & \\[9pt] \hline
 $\hat{\alpha}_{23}$ & 0.210 \newline (-0.10, 0.58) & 0.214 \newline (-0.19, 0.63) & 0.202 \newline (-0.12, 0.50) & 0.196 \newline (-0.06, 0.52) & 0.201 \newline (-0.04, 0.49) & \\[9pt] \hline
 $\hat{\gamma}_E$ & 512 \newline (382, 698) & 511 \newline (413, 620) & 503 \newline (430, 576) & 500 \newline (453, 565) & 499 \newline (457, 552) & \\[9pt] \hline
 $\hat{\gamma}_U$ & 1013 \newline (886, 1169) & 1007 \newline (889, 1139) & 1005 \newline (897, 1159) & 998 \newline (898, 1112) & 1002 \newline (902, 1131) & \\[9pt] \hline
 $\hat{\theta}$ &  -1.010 \newline (-1.33, -0.71) & -0.521 \newline (-0.76, -0.29) & 0.006 \newline (-0.18, 0.18) & 0.504 \newline (0.34, 0.66) & 0.999 \newline (0.81, 1.19) & \\[9pt] \hline
\end{tabular}
\end{table}

We now look at the data generated in Section \ref{effect_differential_ascertainment}, and see if our model is able to detect the presence of differential ascertainment. We focus on the three-list setting, the one relevant for our data analysis in Section \ref{data_analysis}. The parameters used are the same as before. 
Figure \ref{confidence_bands_model_ratio} shows the point estimate and $95\%$ confidence bands for the estimated ratio $\hat{\gamma}_E/\hat{\gamma}_U$ and for the estimated parameter $\hat{\theta}$ when the differential ascertainment is $\theta \in [-1, 1]$. Compared to Figure \ref{confidence_bands_obs_ratio}, we see that the estimated ratio is always centered around the true ratio $0.5$, and that the width of the confidence bands is not extremely sensitive to the value of $\theta$. In the right part of the plot, instead, we see that the estimate $\hat{\theta}$ is correctly centered around the true value of $\theta$, and that again the confidence bands are not very wide. 
Figure \ref{confidence_bands_model_ratio} also allows us to see that, if we use the estimated ratio to test for the presence of differential ascertainment, we would reject the null when the $95\%$ confidence bands do not cover the average observed value, which happens when $|\theta| \gtrsim 0.4$. When using the estimated parameter $\hat{\theta}$, instead, the rejection happens already for $|\theta| \gtrsim 0.25$. This means that, when we are interested in knowing if differential ascertainment is present in our data, it is better to use the estimated $\hat{\theta}$ directly instead of the ratio of exposed and unexposed cases.
In Table \ref{estimate_parameters_simulation} we see the average estimates for all the parameters of the model out of $200$ simulations, together with a $95\%$ confidence interval.

\begin{figure}[!htbp]
\caption{The point estimate and $95\%$ confidence bands for the ratio $\hat{N}^E/\hat{N}^U$ obtained with the saturated log-linear model with 3 lists, when the differential ascertainment parameter is set to be $\theta \in [-1,1]$. The blue line is the observed mean ratio. The data generation process is explained in Section \ref{effect_differential_ascertainment}.}
\label{confidence_bands_CR}
\vspace{5mm}
\centering
\includegraphics[width = 0.65\linewidth]{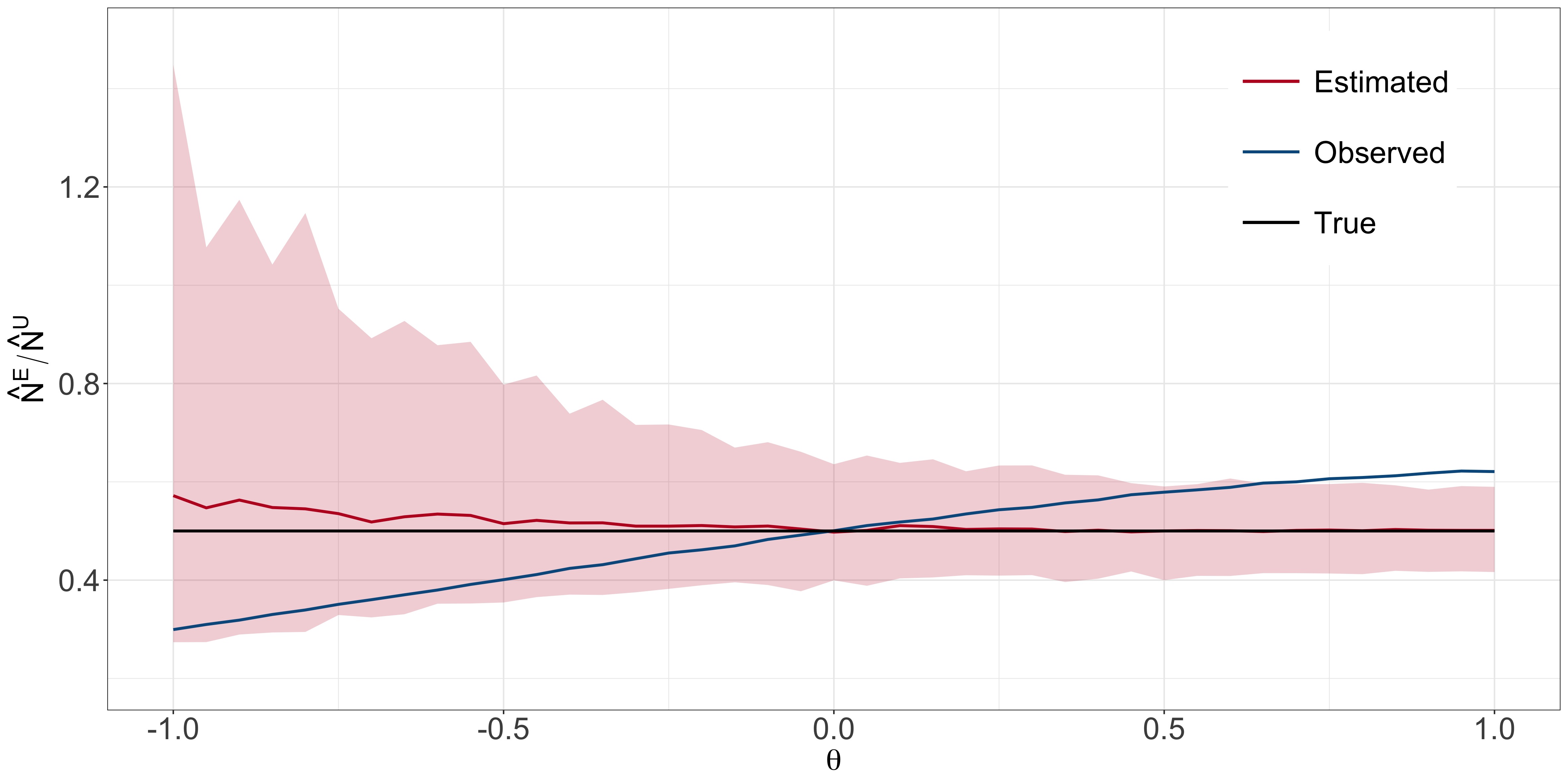}
\end{figure}

One can think that a simpler alternative to this method is to directly estimate $N^E$ and $N^U$, for example using a log-linear model. Using a bootstrap procedure, similar to what we explained at the end of Section \ref{list_dependence}, we can then obtain a confidence interval for the ratio $N^E/N^U$ to attach to the point estimate. The literature on this topic is very rich. For a detailed explanation, see \citet{bfh75} Moreover, \citet{p00} defines a unified log-linear framework allowing for unequal capture probabilities among the population (see also \citealp{c87, pnbh90, ca99}) and \citet{ts99} include the use of covariates in log-linear models.
This procedure, however, even though it is valid in providing a reasonable confidence interval for the ratio of interest, does not yield directly reliable statistical evidence on the presence or not of differential ascertainment. If we use a saturated log-linear model on the data generated in Section \ref{effect_differential_ascertainment}, we can see in Figure \ref{confidence_bands_CR} that the width of the confidence bands heavily depends on the differential ascertainment parameter. Moreover we notice that often, even when differential ascertainment is present, the confidence interval obtained in this way will cover the observed value $N^E_{obs}/N^U_{obs}$, so that we cannot reject the null hypothesis that no differential ascertainment is present.
Even though this procedure is not able to tell us if we are facing differential ascertainment, we can still use it to first estimate the count of individuals in the missing cells, and then apply the complete model in (\ref{complete_model}). 
We implement this idea in the Appendix, and see that the results are consistent with the ones obtained with our method.

\section{Description of the NVDRS Dataset}

To address our question about differential ascertainment of death due to child maltreatment by race, we used a dataset obtained from the National Violent Death Reporting System (NVDRS). Here we provide a brief description of the NVDRS, but a more detailed explanation can be found in \citet{bfjc16}
The NVDRS is a surveillance system created in 2002 from a pilot program called the National Violent Injury Statistics System. In 2010, 19 States were participating in the data collection; 14 additional States joined the system in 2014. The goal of the system is to monitor the occurrence of homicides, suicides, unintentional firearm deaths, deaths of undetermined intent, and deaths from legal intervention (excluding legal executions) in the US. 
The system uses different data sources and merges them together to obtain a more complete description of the event. The primary data sources used for the system are death certificates, coroner/medical examiner reports (including toxicology reports) and law enforcement reports. Coroner/medical examiner reports and law enforcement reports contain narratives that can provide detailed information regarding the circumstances and characteristics of incidents. The manner of death is captured by International Classification of Diseases, Tenth Revision (ICD-10) codes. Other relevant information collected by the system are {demographics (for victims and suspects), method of injury (eg, sharp instrument), place of injury, information on the victim's life stressors, the victim-suspect relationship (eg, spouse or partner), the presence of intimate partner violence (IPV), toxicology (for victims), weapon information and whether other crimes occurred that are related to the incident (eg, robbery followed by homicide).}

As already mentioned, we focus our attention on the deaths of children aged 10 or less, and consider as cases the deaths that were caused by child maltreatment. We are interested in determining if race played a role in the ascertainment of the cases. The timeframe of interest is 2010 to 2015. We follow \citet{lpmsa08} (pg. 11) in considering child maltreatment to be ``any act or series of acts of commission or omission by a parent or other caregiver that results in harm, potential for harm, or threat of harm to a child.''

The dataset consists of 1983 deaths of children from ages 0 to 10 years old and 62 variables that describe the deaths.
Among these variables, there is one that describes the police report, one for the coroner report and others linked to the death certificate, for example the code for cause of death, the manner of death and some descriptive variables. From these variables, we constructed three separate lists, all of them trying to count the total number of deaths caused by child maltreatment. The death certificate (DC) combines information from the ICD-10 codes, the death causes and the caregiver being identified as a suspect.
For the police report (LE), and coroner report (CME), there are in the dataset two variables that define which deaths were considered to be caused by child maltreatment according to the police report and coroner's report respectively. Unfortunately, these variables cannot be trusted to capture all of the cases. A further analysis, described in the Appendix, was needed to incorporate all the cases that were described in the narratives but not automatically reported in the lists.

We want to point out that we are not the first ones to propose a study related to case-controls on the NVDRS data. \citet{femsui06} analysed the risk factors for femicide-suicide in a case-control study that made use of data from $11$ cities in the NVDRS dataset.

\section{Analysis of the NVDRS Data}{\label{data_analysis}}

After the manual scrutiny of the three lists, the total number of cases observed is 968. As anticipated, we define the exposure based on the variable race, and consider exposed the cases where the victim is white and unexposed the cases where the victim is black. The observed ratio of white and black cases is $N^E_{obs}/N^U_{obs} = 539/429 = 1.256$. Table \ref{observed_cases} is the contingency table for the observed cases, divided by whites and blacks (written in boldface). We first want to test if there is evidence that the individual ascertainment parameters $\theta_j$ are list specific, or if instead we can simply assume that there exist a single parameter $\theta$, possibly responsible for differential ascertainment.
\begin{table}[!htbp]
\centering
\caption{Contingency table of exposed (white) and \textbf{unexposed} (black) cases. The cell at the bottom right is missing because it corresponds to the count of individuals that are not ascertained in any of the three lists.}
\label{observed_cases}
\vspace{5mm}
\begin{tabular}{M{1cm} M{0.5cm} M{1.5cm} M{1.5cm} M{1.5cm} M{1.5cm} @{}m{0pt}@{}}
\cline{3-6}
 & \multicolumn{1}{c}{} & \multicolumn{4}{c}{DC} & \\[5pt] \cline{3-6} 
 &  & \multicolumn{2}{c}{Y} & \multicolumn{2}{c}{N} & \\[5pt] \cline{3-6} 
 &  & \multicolumn{4}{c}{LE} & \\[5pt] \cline{3-6} 
\multicolumn{1}{c}{} &  & Y & N & Y & N & \\[5pt] \cline{1-6} \\[-10pt]
\multicolumn{1}{c}{\multirow{2}{*}{CME}} & \multicolumn{1}{c}{Y} & $207, \textbf{166}$ & $53, \textbf{35}$ & $139, \textbf{118}$ & $58, \textbf{35}$ & \\[5pt] \cline{2-6} \\[-10pt]
\multicolumn{1}{c}{} & \multicolumn{1}{c}{N} & $23, \textbf{24}$ & $15, \textbf{11}$ & $44, \textbf{40}$ & ? & \\[5pt] \cline{1-6} 
\end{tabular}
\end{table}
To do so, we fit the incomplete model (\ref{incomplete_model}) to these data using the probabilities in (\ref{prob_rasch_no_indep}). The order of the lists is the one presented in Table \ref{contingency_table_3lists}, so DC represents list 1, LE list 2 and CME list 3. With this notation, we get the parameter estimates in Table \ref{estimates_diff_theta} and a value for the log-likelihood of $\ell = 3457.38$.
\begin{table}[!htbp]
\centering
\caption{MLE estimates of the parameters using the incomplete data in Table \ref{observed_cases}, and allowing a different capture propensity parameter $\theta_j$ for each list.}
\label{estimates_diff_theta}
\vspace{2mm}
\begin{tabular}{M{.85cm} M{.85cm} M{.85cm} M{.85cm} M{.85cm} M{.85cm} M{.7cm} M{.7cm} M{1cm} M{1cm} M{.85cm} @{}m{0pt}@{}}
\hline \\[-10pt]
$\hat{\alpha}_1$  & $\hat{\alpha}_2$ & $\hat{\alpha}_3$ & $\hat{\alpha}_{12}$ & $\hat{\alpha}_{13}$ & $\hat{\alpha}_{23}$ & $\hat{\gamma}_E$ & $\hat{\gamma}_U$ & $\hat{\theta}_1$ & $\hat{\theta}_2$ & $\hat{\theta}_3$ & \\[5pt] \hline \\[-10pt]
0.056  & 0.867 & 0.175 & 0.575 & 0.954 & 0.863 & 580 & 459 & -0.002 & -0.241 & 0.152 & \\[5pt] \hline
\end{tabular}
\end{table}
There seems to be a fairly large difference in the $\theta_j$'s, especially between $\theta_2$ and $\theta_3$. To test if this difference is significant, we fit on the same data the constrained model, which assumes $H_0: \theta_1 = \theta_2 = \theta_3$. Here, we get the parameter estimates in Table \ref{estimates_single_theta}, and a value for the log-likelihood of $\ell_0 = 3455.60$. The generalised likelihood ratio statistic is then $\Lambda = 2\cdot (3457.38 - 3455.60) = 3.56$, which has p-value of $0.169$ on a $\chi^2$ distribution with $2$ degrees of freedom. Given this result, we cannot reject the null hypothesis that all the $\theta_j$'s have the same value. We will then continue our analysis assuming from now on that there is only one parameter $\theta$. We want to test if there is any evidence of differential ascertainment, i.e., if such $\theta$ is significantly different from $0$, and if the estimated ratio is different from the observed one.

\begin{table}[!htbp]
\centering
\caption{MLE estimates of the parameters using the incomplete data in Table \ref{observed_cases}, fitting the model in (\ref{incomplete_model}) under the null hypothesis $H_0: \theta_1 = \theta_2 = \theta_3 = \theta$.}
\label{estimates_single_theta}
\vspace{2mm}
\begin{tabular}{M{1.1cm} M{1.1cm} M{1.1cm} M{1.1cm} M{1.1cm} M{1.1cm} M{.9cm} M{.9cm} M{1.1cm} @{}m{0pt}@{}}
\hline \\[-10pt]
$\hat{\alpha}_1$  & $\hat{\alpha}_2$ & $\hat{\alpha}_3$ & $\hat{\alpha}_{12}$ & $\hat{\alpha}_{13}$ & $\hat{\alpha}_{23}$ & $\hat{\gamma}_E$ & $\hat{\gamma}_U$ & $\hat{\theta}$ & \\[5pt] \hline \\[-10pt]
0.074  & 0.750 & 0.289 & 0.572 & 0.951 & 0.848 & 580 & 459 & -0.033 & \\[5pt] \hline 
\end{tabular}
\end{table}

The estimated values $\hat{\gamma}_E$ and $\hat{\gamma}_U$ in Table \ref{estimates_single_theta} tell us that the unobserved values in the missing cell are $580 - 539 = 41$ white cases and $459 - 429 = 30$ black cases. Our estimate of the true underlying ratio of exposed and unexposed cases is $\hat{\gamma}_E/\hat{\gamma}_U = 1.264$, which is close to the observed one (yet slightly bigger). This is consistent with the fact that $\hat{\theta}$ is very close to $0$ and negative, so we get the result anticipated in Remark \ref{why_theta}.

\begin{figure}[!htbp]
   \caption{Histogram of $\hat{\theta}$ under the null $H_0:\theta = 0$ (left) and histogram of $\hat{\gamma}_E/\hat{\gamma}_U$ under the null $H_0: \gamma_E/\gamma_U = 1.256$ (right) obtained generating $1000$ new populations with the procedures described in Section \ref{list_dependence}. The blue solid line is the value estimated from the data, the dash-dotted lines are the $0.05$ and $0.95$ quantiles and the dashed lines are the $0.025$ and $0.975$ quantiles.}
   \label{empirical_distribution_thetahat}
   \begin{minipage}{0.5\textwidth}
     \centering
     \includegraphics[width=.9\linewidth]{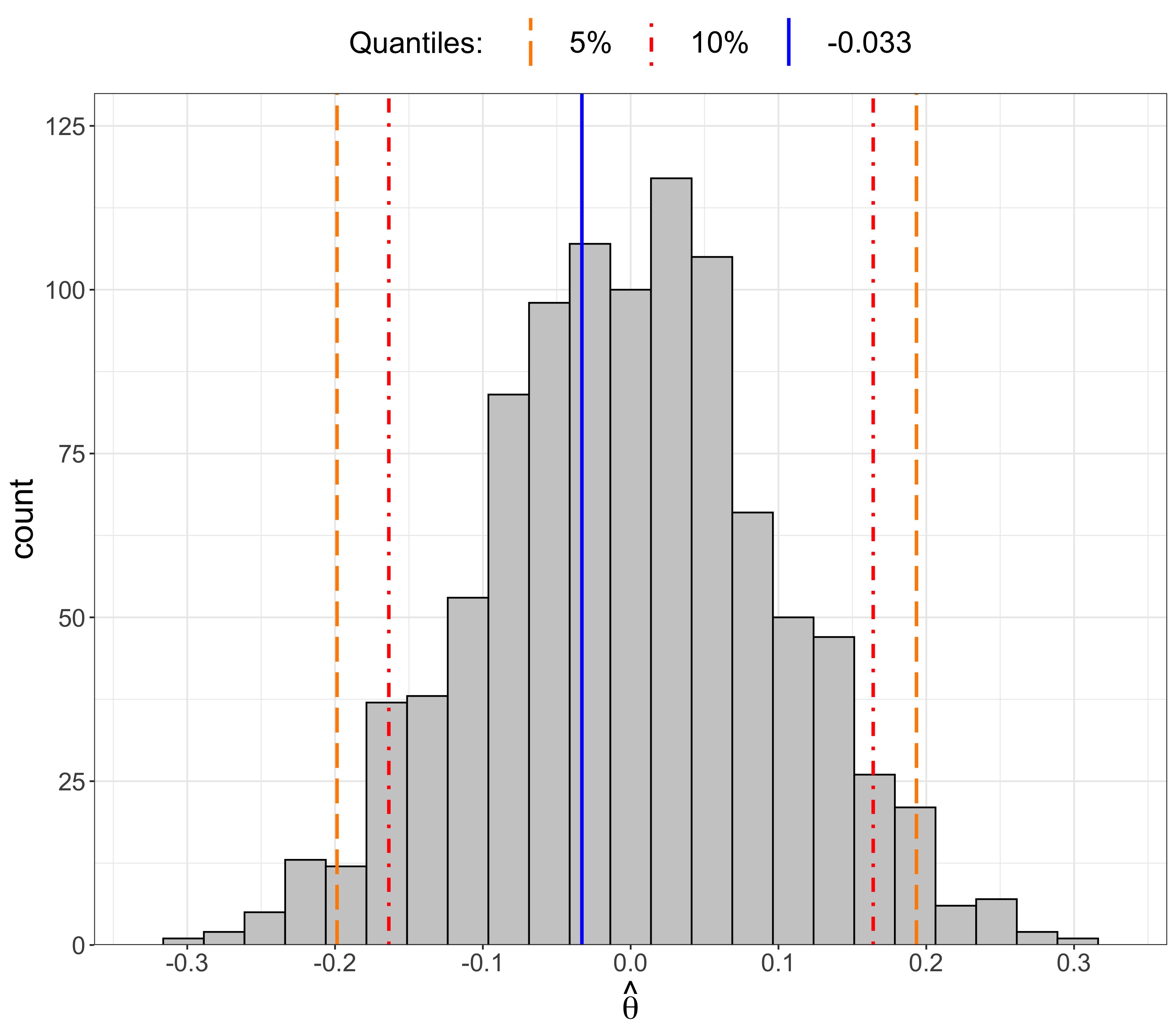}
   \end{minipage}\hfill
   \begin{minipage}{0.5\textwidth}
     \centering
     \includegraphics[width=.9\linewidth]{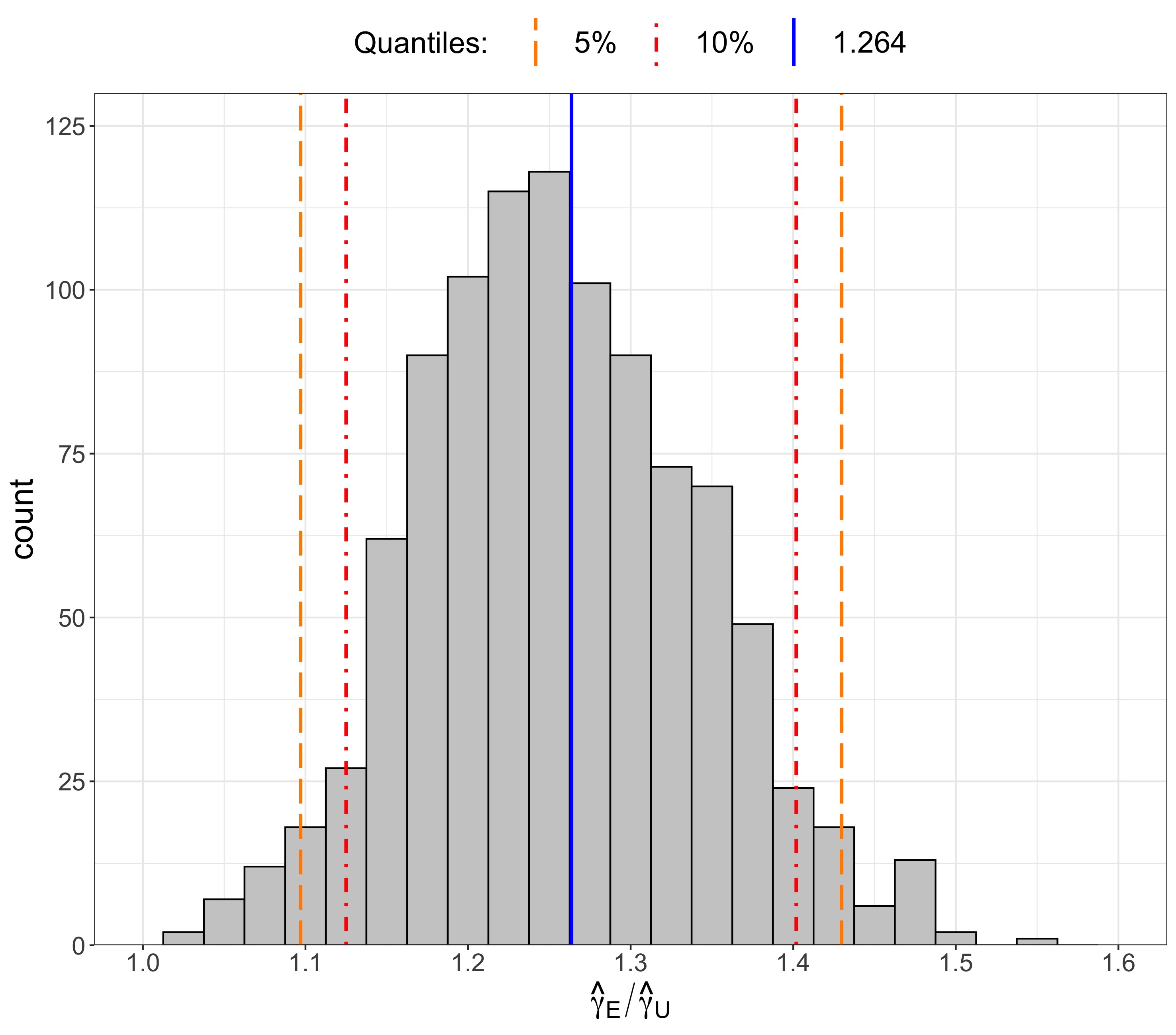}
   \end{minipage}
\end{figure}

A confidence interval for $\hat{\gamma}_E/\hat{\gamma}_U$ under the null hypothesis $H_0: \gamma_E/\gamma_U = 1.256$ can be obtained in the way explained is Section \ref{list_dependence}, by first fitting the incomplete model (\ref{incomplete_model}) under the constraint $\gamma_E = 1.256\cdot \gamma_U$ (the maximum likelihood estimators are reported in Table \ref{estimates_no_theta}) and then creating new populations. When fitting the model on these new populations, we get a $95\%$ confidence interval of $[1.097, 1.430]$, which means that $H_0$ is not rejected. The empirical distribution for the ratio is reported in the right panel of Figure \ref{empirical_distribution_thetahat}, together with the estimated ratio.

\begin{table}[!htbp]
\centering
\caption{MLE estimates of the parameters using the incomplete data in Table \ref{observed_cases}. In the first row we used the null hypothesis $H_0: \theta = 0$, while in the second row we used $H_0: \gamma_E/\gamma_U = 1.256$ so $\hat{\gamma}_E$ is not directly estimated bu just set to be $1.256\cdot\hat{\gamma}_U$.}
\label{estimates_no_theta}
\vspace{2mm}
\begin{tabular}{M{1.1cm} M{1.1cm} M{1.1cm} M{1.1cm} M{1.1cm} M{1.1cm} M{1cm} M{1cm} M{1.2cm} @{}m{0pt}@{}}
\hline \\[-10pt]
$\hat{\alpha}_1$  & $\hat{\alpha}_2$ & $\hat{\alpha}_3$ & $\hat{\alpha}_{12}$ & $\hat{\alpha}_{13}$ & $\hat{\alpha}_{23}$ & $\hat{\gamma}_E$ & $\hat{\gamma}_U$ & $\hat{\theta}$  \\[5pt] \hline \\[-10pt]
0.055  & 0.730 & 0.266 & 0.574 & 0.953 & 0.852 & 579 & 461 & & \\[5pt] \hline \\[-10pt]
0.055  & 0.730 & 0.266 & 0.574 & 0.953 & 0.852 &  & 460 & 0.004 & \\[5pt] \hline
\end{tabular}
\end{table}
In the left panel of Figure \ref{empirical_distribution_thetahat}, instead, we have the empirical distribution of $\hat{\theta}$ under the null hypothesis $H_0 : \theta = 0$. It is obtained fitting the model on the NVDRS data using the probabilities $p^U_{xyz}$ in (\ref{prob_rasch_no_indep}) for both exposed and unexposed cases (the estimates are reported in Table \ref{estimates_no_theta}). Using those values, we generate new incomplete populations, and fit the incomplete model to obtain a $95\%$ confidence interval for $\theta$ under the null of $[-0.199, 0.193]$. Notice that, keeping all other parameters fixed, this range of values for $\theta$ can make the probability $p_{000}^E$ range in $[0.046, 0.093]$.

From this analysis we can conclude that there is no evidence of differential ascertainment by race in the NVDRS data that we analysed.

\section{Discussion}
In this work, we developed a method to test for the presence of differential ascertainment when data are collected in multiple lists. We showed through a simulation the dangerous effect that differential ascertainment can have on estimating the odds-ratio in observational studies, and used the proposed method to detect when it is present and correct its effect.
We applied it to a dataset obtained from the National Violent Death Reporting System, where the cases of child maltreatment that led to death of children from 0 to 10 years old are recorded. When considering race as an exposure, the observed ratio of the cases is $1.256$. We showed that the estimated ratio is very close to the observed one, and used the confidence intervals for $\hat{\theta}$ and for $\hat{\gamma}_E/\hat{\gamma}_U$ to test whether there is any significant presence of differential ascertainment. 

In our analysis we found no evidence of differential ascertainment by race in the NVDRS data. The point estimate for the differential ascertainment parameter $\theta$ is $-0.033$ which is contained in the $95\%$ confidence interval for the parameter under the null $\theta=0$. 
The confidence interval tells us that plausible values for the amount of differential ascertainment are approximately $-0.2$ to $0.2$. In the extreme cases, the probability of not ascertainment of an element with exposure in each of the three list can go from $0.046$ to $0.093$.
This tells us that, although we cannot guarantee that differential ascertainment is not present in our data, we have no evidence of a significant impact.
Notice that in this work we focused on the situation in which only the cases can be affected by differential ascertainment. However, this model can also be used in the situation where the controls might suffer from the same complication. In that case, one needs to keep into consideration the randomness in the estimate of the ratios for both cases and controls, to get a confidence interval for the odds ratio (\ref{or_equation}).

Apart from the direct application of the method on multiple lists from the whole population of interest, one can think about using it as an exploratory analysis on a smaller sample from the whole population, when it may be expensive to collect data from multiple lists rather than just one. In this case, an initial test for differential ascertainment in the small sample can give an idea of the possible bias that we would get if we were to only ascertain the data using a single list.

\section*{Acknowledgements}

The authors are grateful to Dr. Joanne Klevens for useful discussions and suggestions on the NVDRS data.

\clearpage

\bibliographystyle{apalike}

\newpage
\appendix

\section{Modification of the NVDRS dataset}
Reading the police and coroner narratives, it gets clear that the completeness of the variables LE\_DeathAbuse and CME\_DeathAbuse cannot be trusted. In multiple circumstances the narrative explicitly classifies an individual as a victim of child abuse or maltreatment, while the corresponding variable does not report it as such.
For this reason we had to manually check all the narratives of the individuals that were not initially classified as cases. Below are the details of how the cases were defined in each of the three lists.
\begin{itemize}
\item[DC:] For creating the death certificate list, we didn't have access to a full narrative, but only to a group of variables. Following a procedure similar to the one described in \citet{kl10} we coded a death as being caused by child maltreatment if (1) the ICD-10 code was Y06 or Y07 (where Y06 refers to neglect and abandonment and Y07 to other maltreatment syndromes), (2) the ICD-10 code was one of assault, i.e., X85, X86, X87, X88, X89, X90, X91, X92, X93, X94, X95, X96, X97, X98, X99, Y00, Y01, Y02, Y03, Y04, Y08 or Y09 and the variable DeathCause contained one of the following words: \textit{abandon, head trauma, shaken, abusive} or \textit{abuse}, and (3) the ICD-10 code was one of assault and the suspect was the caregiver.

This list captured 298 white cases and 236 black cases, for a total of 534.

\item[LE:] We started considering as cases the individuals for which LE\_DeathAbuse = 1. Moreover we also read the narratives of the victims that were initially coded as controls, to account for the inefficiency of the variable LE\_DeathAbuse. Here we followed the definition of abuse contained in \citet{lpmsa08} and in particular we focused on the following aspects:
\begin{itemize}
\item The perpetrator has to be a caregiver of the victim, i.e., a person that ``at the time of the maltreatment is in a permanent or temporary custodial role.''
\item When considering acts of commission (Child Abuse) the action should be intentional, but the harm to the child need not be the intended consequence. The acts of omission (Child Neglect) are more rare because NVDRS tends to not capture them, but also in this case the harm to the child need not be the intended consequence.
\end{itemize}

Initially there were 495 cases captured by the variable LE\_DeathAbuse regarding individuals of race black or white. We manually added another 266, bringing the total number to 761. Of these, 413 are white and 348 are black.

\item[CME:] We used the same rule as for the police narrative, initially using the variable CME\_DeathAbuse and then reading the remaining narratives.

Here we initially had 534 cases captured by the variable CME\_DeathAbuse, and we manually added another 277. Of the 811 total cases, 457 are white and 354 are black.
\end{itemize}

\section{Data Analysis Estimating the Missing Counts First}{\label{capture_recapture}}

As anticipated in Section \ref{model_performance}, we can use a log-linear model to first estimate the total counts $N^E$ and $N^U$, and then apply the complete model in (\ref{complete_model}). 
We decided to adopt the saturated model, to impose no constraints on the parameters. It is also possible here to use the model selection procedure proposed in \citet{bbsda99}, in case one wants to employ a more parsimonious model. The number of unobserved exposed cases is estimated to be $47$, while for the unobserved unexposed cases is $26$. The estimated ratio is then $586/455 = 1.288$.
To get the $95\%$ confidence interval for the ratio $N^E/N^U$ using this log-linear model, we resample the observed data using a multinomial distribution with the empirical probability for each cell. Then estimate the count of the missing cells using the saturated (or selected) log-linear model. The confidence interval that we would obtain is $[1.16, 1.43]$.

\begin{figure}[!htbp]
\caption{Histogram of $\hat{\theta}$ obtained generating $1000$ new populations and using capture-recapture with model selection. The solid line is the observed value $\hat{\theta} = -0.092$, the dashed lines are the $0.05$ and $0.95$ quantiles and the dash-dotted lines are the $0.025$ and $0.975$ quantiles.}
\label{empirical_distribution_thetahat_capture_recapture}
\vspace{5mm}
\centering
\includegraphics[width=0.6\columnwidth]{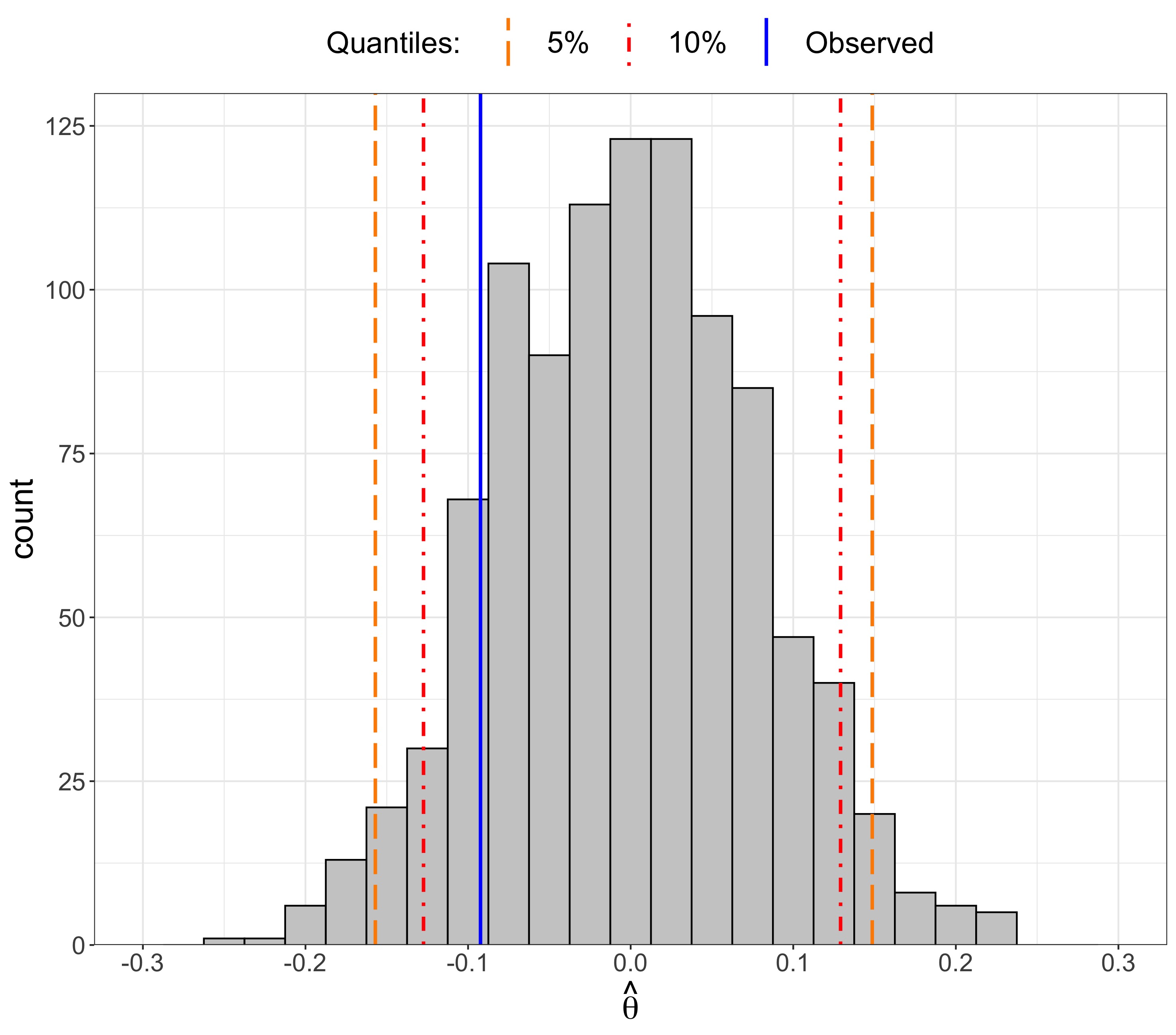}
\end{figure}

If we want to also study the presence of differential ascertainment, instead, we can assume that the total counts of cases is $N^E = 586$ exposed and $N^U = 455$ unexposed, which will be considered fixed. 
We can now fit the complete model (\ref{complete_model}), in which the only source of randomness comes from the multinomial distributions $M^e | N^e$ of the cases in the two contingency tables. Again we expect the estimate for $\theta$ to be negative and small in magnitude, as per Remark \ref{why_theta}.
The estimates that we get from this model are reported in the first line of Table \ref{estimates_model_1}.
\begin{table}[!htbp]
\centering
\caption{MLE estimates of the parameters in the model. We used a saturated log-linear model to estimate the count of the missing cells. In the second line the estimates are under the null $\theta = 0$.}
\label{estimates_model_1}
\vspace{5mm}
\begin{tabular}{M{1.2cm} M{1.2cm} M{1.2cm} M{1.2cm} M{1.2cm} M{1.2cm} M{1.2cm} @{}m{0pt}@{}}
\hline \\[-10pt]
$\hat{\alpha}_1$ & $\hat{\alpha}_2$ & $\hat{\alpha}_3$ & $\hat{\alpha}_{12}$ & $\hat{\alpha}_{13}$ & $\hat{\alpha}_{23}$ & $\hat{\theta}$ & \\[5pt] \hline
0.104 & 0.773 & 0.305 & 0.583 & 0.961 & 0.861 & -0.092 & \\[5pt] \hline
0.052 & 0.720 & 0.247 & 0.584 & 0.961 & 0.868& & \\[5pt] \hline
\end{tabular}
\end{table}
\noindent
We now want to find the empirical distribution of $\hat{\theta}$ under $H: \theta = 0$.
We start by fitting the model with $\theta = 0$ on the complete tables, with the numbers $47$ and $26$ in the bottom right cells of the exposed and unexposed contingency tables respectively. The estimates we got are reported in the second line of Table \ref{estimates_model_1}. 
Using them, we generate $1000$ new populations from two multinomial distributions with probabilities given by (\ref{prob_rasch_no_indep}). 
We then fit the complete model on these populations, and get the empirical distribution for $\hat{\theta}$ which is plotted in Figure \ref{empirical_distribution_thetahat_capture_recapture}, together with the quantiles that define a $90\%$ and $95\%$ confidence interval. Let $\alpha = 0.05$, the relevant quantiles are $q_{\alpha/2} = -0.157 , \ q_{\alpha} = -0.127 , \ q_{1-\alpha} = 0.129 $ and $q_{1-\alpha/2} = 0.149$.

\section{An Extension of the Model}{\label{extension}}

As anticipated in Section \ref{model}, we can consider an extension of the model that keeps into account the individual differences in the propensity to be captured. Instead of simply setting $\theta_i^U = 0$ and $\theta_j^E = \theta$ for any $i = 1,..., N^E$, $j = 1,..., N^U$, we allow these values to vary, and assume they are normally distributed, $\theta_i^U \sim N(0, \sigma^2)$ and $\theta_i^E \sim N(\mu, \sigma^2)$. Each value is generated independently, and the parameter $\mu$ now represents the mean amount of differential ascertainment. 

We start by examining the situation in which the two contingency tables are complete. In our main example on NVDRS data, this means that we apply the log-linear model described before, and add the values 85 and 63 in the empty cells of the exposed and unexposed contingency tables respectively. We can consider the total counts $N^e$ as fixed, and the likelihood of each contingency table is then:
\begin{align}{\label{likelihood_integral_complete}}
\L(M^e | N^e) = \int_{-\infty}^\infty \L(M^e | N^e, \theta) \cdot \phi_e(\theta) \ d\theta
\end{align}
Here $\L(M^e | N^e, \theta)$ is the multinomial likelihood where the total count is $N^e$ and the probabilities are given in (\ref{prob_rasch_no_indep}). The density $\phi_e(\theta)$ is the normal density with mean $\mu$ for the exposed, and $0$ for the unexposed cases. When fitting this likelihood on the NVDRS data, again using Assumption \ref{individual_independence}, we get the estimates reported in the first row of Table \ref{estimates_model_3_4}.
\begin{table}[!htbp]
\centering
\caption{MLE estimates of the parameters in the extended model. In the first row we have the estimates of the parameters when using capture-recapture to first estimate the population size (and it is very similar to the first row in Table \ref{estimates_model_1}). In the second row we use the Poisson assumption on the total count of cases, and the estimates are similar to the ones in Table \ref{estimates_single_theta}.}
\label{estimates_model_3_4}
\vspace{5mm}
\begin{tabular}{M{1.1cm} M{1cm} M{1.1cm} M{1cm} M{1cm} M{1cm} M{0.8cm} M{0.8cm} M{1.1cm} M{1.1cm} @{}m{0pt}@{}}
\hline
$\hat{\alpha}_1$ & $\hat{\alpha}_2$ & $\hat{\alpha}_3$ &$\hat{\alpha}_{12}$ &$\hat{\alpha}_{13}$ &$\hat{\alpha}_{23}$ & $\hat{\gamma}_W$ & $\hat{\gamma}_B$  &$\hat{\mu}$     & $\hat{\sigma}$ & \\[5pt] \hline
0.118&0.721&0.345&0.733&0.971&0.819 & & & -0.110&0.0009&  \\[5pt] \hline
0.074&0.750&0.288&0.572&0.952&0.848&580&459&-0.033&0.0043 & \\[5pt] \hline
\end{tabular}
\end{table}
We happily notice that these values are extremely similar to the ones in Table \ref{estimates_model_1}, and that $\hat{\sigma}$ is very small. This suggests that the original model with $\theta_i^U = 0$ and $\theta_i^E$ fixed is appropriate for the data we are studying. To confirm this theory, we consider the natural extension of our original model, where the contingency tables are incomplete and the individual capture strengths are normally distributed. We assume as before that $N^e \sim \text{Poisson}(\gamma_e)$.
Starting from (\ref{multinomial_likelihood_observed}), and treating the complete contingency table as in (\ref{likelihood_integral_complete}), we have
\begin{equation}
\begin{aligned}
\L(M^e_{obs}) &= \int_{-\infty}^\infty \L(M^e_{obs} | \theta) \cdot \phi_e(\theta) \ d\theta \\
&= \int_{-\infty}^\infty \frac{\prod_{x,y,z} {(p_{xyz}^e)}^{M_{xyz}^e} \cdot \gamma_e^{N_{obs}^e}\cdot e^{\gamma_e (p_{000}^e - 1)}}{\prod_{x,y,z} M_{xyz}^e!} \cdot \phi_e(\theta) \ d\theta
\end{aligned}
\end{equation}
Fitting this likelihood to our data, we get the estimates in the second line of Table \ref{estimates_model_3_4}.
Again it looks like allowing variability in the individual capture strength is not necessary, since the estimates are nearly the same as the ones in Table \ref{estimates_single_theta}, and $\hat{\sigma}$ is very small.

\section{Identifiability of the Parameters}{\label{identifiability}}

Here we discuss the problem of identifiability in our model, and show that when the number of lists is $J \geq 2$ all the parameters are identifiable.
Identifiability means that if we have two distinct set of parameters, then the corresponding probability distributions of our model are different. To prove identifiability one usually assumes that two set of parameters induce the same distribution and proves that this implies that the parameters are actually the same in the two set. We denote the second set of parameters with a prime symbol ', and the probabilities defined with those parameters by ${p'}_{ij}^e$.

The likelihood for each incomplete contingency table when the number of lists is $J = 3$ is in (\ref{extended_likelihood}). When the number of lists is different, only the sum needs to be changed to take into account all the observed entries of the contingency tables.
Considering together exposed and unexposed cases as in (\ref{incomplete_model}), and ignoring the denominator of the likelihood (since it does not depend on the parameters), we get that the log-likelihood of the model is proportional to
\begin{equation}{\label{log_likelihood}}  
\ell(M_{obs}^E, M_{obs}^U) = \sum_{x,y,z}\l( M_{xyz}^E \log(\gamma_E \cdot p_{xyz}^E) +  M_{xyz}^U \log(\gamma_U \cdot p_{xyz}^U)\r) + \gamma_E (p_{000}^E-1) + \gamma_U (p_{000}^U-1)
\end{equation}
When $J = 1$ (there is only one list) we do not have identifiability. In fact for the two sets of parameters $$\l(\theta = 0, \alpha_1 = 1, \gamma_E = \frac{1+e}{2}, \gamma_U = \frac{1+e}{2}\r) \quad\text{and}\quad \l(\theta' = 0, \alpha_1' = 0, \gamma_E' = e, \gamma_U' = e\r)$$ 
the log-likelihood is $(M_1^E+M_1^U) \cdot \log\l(\frac e2\r) - e$, while for 
$$\l(\theta = 1, \alpha_1 = 0, \gamma_E = 1+e, \gamma_U = 2e\r) \quad\text{and}\quad \l(\theta' = 0, \alpha_1' = 1, \gamma_E' = 1+e, \gamma_U' = 1+e\r)$$
the log-likelihood is $M_1^E + M_1^U - 2e$.
This proves that we can get the same likelihood with distinct sets of parameters, hence the model is not identifiable.

When the number of lists is $J = 2$, we show that we already have identifiability. In this case, in fact, we only add the parameter $\alpha_2$ (since the higher order interaction is always excluded). 
We see from (\ref{log_likelihood}) that the conditions for having the same distribution with the two set of parameters is that 
$$\gamma_e \cdot p_{ij}^e = \gamma_e' \cdot {p'}_{ij}^e \qquad\text{for any exposure and any } i, j \ \text{s.t. } i+j > 0 .$$
Here we have that, for the exposed contingency table, the probabilities are
$$p_{11}^E = \frac{e^{\theta+\alpha_1}\cdot e^{\theta + \alpha_2}}{(1+e^{\theta+\alpha_1})(1+e^{\theta+\alpha_2})}, \ p_{10}^E = \frac{e^{\theta+\alpha_1}}{(1+e^{\theta+\alpha_1})(1+e^{\theta+\alpha_2})}, $$
$$p_{01}^E = \frac{e^{\theta+\alpha_2}}{(1+e^{\theta+\alpha_1})(1+e^{\theta+\alpha_2})}, \ p_{00}^E = \frac{1}{(1+e^{\theta+\alpha_1})(1+e^{\theta+\alpha_2})}$$
and same for the unexposed, without the parameter $\theta$. 
If we combine together the equation involving $p_{11}^U$ and the one with $p_{10}^U$, we get that $\alpha_2 = \alpha_2'$. Similarly using the equation with $p_{11}^U$ and the one with $p_{01}^U$ we have $\alpha_1 = \alpha_1'$. It is simple now to use the equations for the exposed cases and show that $\theta = \theta'$ and finally that $\gamma_E = \gamma_E'$ and $\gamma_U = \gamma_U'$. This proves that the model is identifiable.

When the number of lists is larger, we have $2^J - 1$ expressions $\gamma_U \cdot p_{i_1,..., i_J}^U = \gamma_U' \cdot {p'}_{i_1,..., i_J}^U$, from which we get identifiability of the $2^J - 2$ lists parameters ($J$ for the individual lists, $\binom{J}{2}$ for the two-list interaction and so on). From three of the remaining $2^J - 1$ expressions $\gamma_E \cdot p_{i_1,..., i_J}^E = \gamma_E' \cdot {p'}_{i_1,..., i_J}^E$ we get identifiability of $\theta, \gamma_E$ and $\gamma_U$.

\begin{framed}
\begin{remark}
The same result on identifiability holds if we allow different $\theta_j$ coefficients for each list. In the situation with only $2$ lists we now have $6$ observations and $6$ parameters, and similarly as before it is possible to get identifiability of $(\alpha_1, \alpha_2, \theta_1, \theta_2, \gamma_E, \gamma_U)$. When we extend this result to $J > 2$ lists we can use the $2^J - 1$ expressions $\gamma_U \cdot p_{i_1,..., i_J}^U = \gamma_U' \cdot {p'}_{i_1,..., i_J}^U$ to we get identifiability of the $\alpha$ parameters and the remaining $2^J - 1$ expressions $\gamma_E \cdot p_{i_1,..., i_J}^E = \gamma_E' \cdot {p'}_{i_1,..., i_J}^E$ to solve for $\theta_1,..., \theta_J, \gamma_E$ and $\gamma_U$.
\end{remark}
\end{framed}

\end{document}